\documentclass[a4paper]{article}
\usepackage{INTERSPEECH2022,amssymb,amsmath,epsfig}
\usepackage{caption}
\usepackage{subcaption}
\usepackage{multirow}
\usepackage{dirtytalk}

\usepackage{INTERSPEECH2022}

\ninept
\def\reg{{\rm\ooalign{\hfil
     \raise.07ex\hbox{\scriptsize R}\hfil\crcr\mathhexbox20D}}}

\usepackage{color}
\usepackage[dvipsnames]{xcolor}
\usepackage{soul}

\def\reg{{\rm\ooalign{\hfil
     \raise.07ex\hbox{\scriptsize R}\hfil\crcr\mathhexbox20D}}}

\ninept
\begin{document}
\title{Towards Disentangled Speech Representations}
\name{Cal Peyser$^{12}$, Ronny Huang$^2$, Andrew Rosenberg$^2$, Tara N. Sainath$^2$, Michael Picheny$^1$, Kyunghyun Cho$^1$}
\address{
    $^1$Center for Data Science, New York University, New York City, USA \\
    $^2$Google Inc., U.S.A}
\email{cpeyser@google.com}

\maketitle
\begin{abstract}
The careful construction of audio representations has become a dominant feature in the design of approaches to many speech tasks.  Increasingly, such approaches have emphasized \say{disentanglement}, where a representation contains only parts of the speech signal relevant to transcription while discarding irrelevant information.  In this paper, we construct a representation learning task based on joint modeling of ASR and TTS, and seek to learn a representation of audio that disentangles that part of the speech signal that is relevant to transcription from that part which is not.  We present empirical evidence that successfully finding such a representation is tied to the randomness inherent in training.  We then make the observation that these desired, disentangled solutions to the optimization problem possess unique statistical properties.  Finally, we show that enforcing these properties during training improves WER by 24.5\% relative on average for our joint modeling task.  These observations motivate a novel approach to learning effective audio representations.
\end{abstract}

\section{Introduction \label{sec:introduction}}

The recent success of unsupervised pretraining in language processing can be credited to the advent of sophisticated techniques for learning representations of text.  Methods like ELMo \cite{Peters2018}, GPT \cite{Radford2018}, and BERT \cite{Devlin2018} work by using an unsupervised task that develops a representation of text that is useful for downstream tasks in a way that is agnostic to what that task is.      

In applying these lessons to unsupervised pretraining in speech great progress has been made with the discovery that in a data-intensive domain like audio, it is best to learn a representation that discards unimportant parts of the signal. Contrastive estimation \cite{Gutmann2010}, in which a full reconstruction is not learned, has yielded representations that achieve strong results in speaker identification and speech recognition \cite{Oord2018}.  State-of-the-art methods combine contrastive learning with masked language modeling as in Wav2Vec 2.0 \cite{ Baevski2020} and Adaptive SpecAugment \cite{zhang20}.

Such successes can be seen as signaling a movement away from task-agnostic representations and towards \say{lossy} representations, in which a model learns not only to summarize relevant portions of a signal but also to discard portions that are irrelevant to the downstream task.  This distinction is particularly clear in the world of multi-modal representation learning, where we seek a representation specifically of the intersection between two domains (e.g. audio and images \cite{Peri2021, Morgado2020} or audio and text \cite{Chung2019UnsupervisedLO}).  However, while there are several natural methods for learning a representation that models components of a signal that are required for a task, it is difficult to craft a method that compels a model to specifically exclude irrelevant components.  Approaches in this space have commonly relied on techniques like adversarial learning to exclude particular parts of a signal thought to be irrelevant, as in \cite{Wang2021adversarially}.

In this study, we present a novel architecture specifically designed to learn a measurably disentangled representation of audio using supervised data.  Our model is based on the paradigm of dual learning \cite{DualLearningBook, Xia2016, Wang2019MultiAgentDL, Wang2018DualTL}, which seeks to exploit the \say{duality} between ASR and TTS.  Traditionally, this is done by training a model that performs both ASR and TTS with a shared encoder that is tasked with representing inputs from both the speech and text domains \cite{ren2020unsupervised, xu2020lrspeech}.  Our model adds a secondary encoder, which is intended to capture specifically those parts of the audio signal that are irrelevant to the transcript.  While the primary encoder is utilized for both ASR and TTS, this secondary encoder is used only for audio reconstruction, which is a task that requires both that part of the audio signal that predicts the transcript and the \say{residual} signal that does not.  We argue that disentanglement is facilitated by the explicit modeling of the residual signal by the secondary encoder, and demonstrate this disentanglement by training a speaker-ID classifier on the outputs of both the primary and secondary encoders.

Other studies have shown that in scenarios where more than one solution to an optimization problem is possible (such as generalized vs. overfit solutions \cite{Weber2018} and selection of significant units in a DNN \cite{Frankle2018}), the stochasticity of parameter initialization and minibatch selection can be decisive.  We present empirical evidence that speech signal disentanglement is such a problem.  We find that both entangled and disentangled solutions to our dual learning problem are possible, and that the superior, disentangled solution is arrived at randomly.  We then observe that the disentangled solution has the unique statistical property of using a large amount of its variational capacity in both encoders.  Finally, we show that enforcing this property during training with an additional loss term substantially improves ASR quality.

Possible applications of our joint modeling task include refinement of back-transcription based semi-supervised learning systems such as speech chains \cite{Tjandra17} and Sequential MixMatch \cite{ZChen21}.  We believe that our discoveries motivate the usage of a secondary encoder in such systems to achieve disentanglement in semi-supervised audio representations.

The rest of this paper is structured as follows.  Section \ref{sec:architecture} describes our architecture for a joint audio and text model that can simultaneously perform ASR, TTS, and audio and text reconstruction. Section \ref{sec:experiments} presents the design for our experiment investigating the nature of disentangled representations.  Section \ref{sec:results} details the results of that experiment and observations of the statistical differences between entangled and disentangled representations.  We summarize our findings and discuss future work in Section \ref{sec:conclusions}.

\section{Architecture \label{sec:architecture}}

\begin{figure*}[t!]
  \centering
  \includegraphics[scale=0.175]{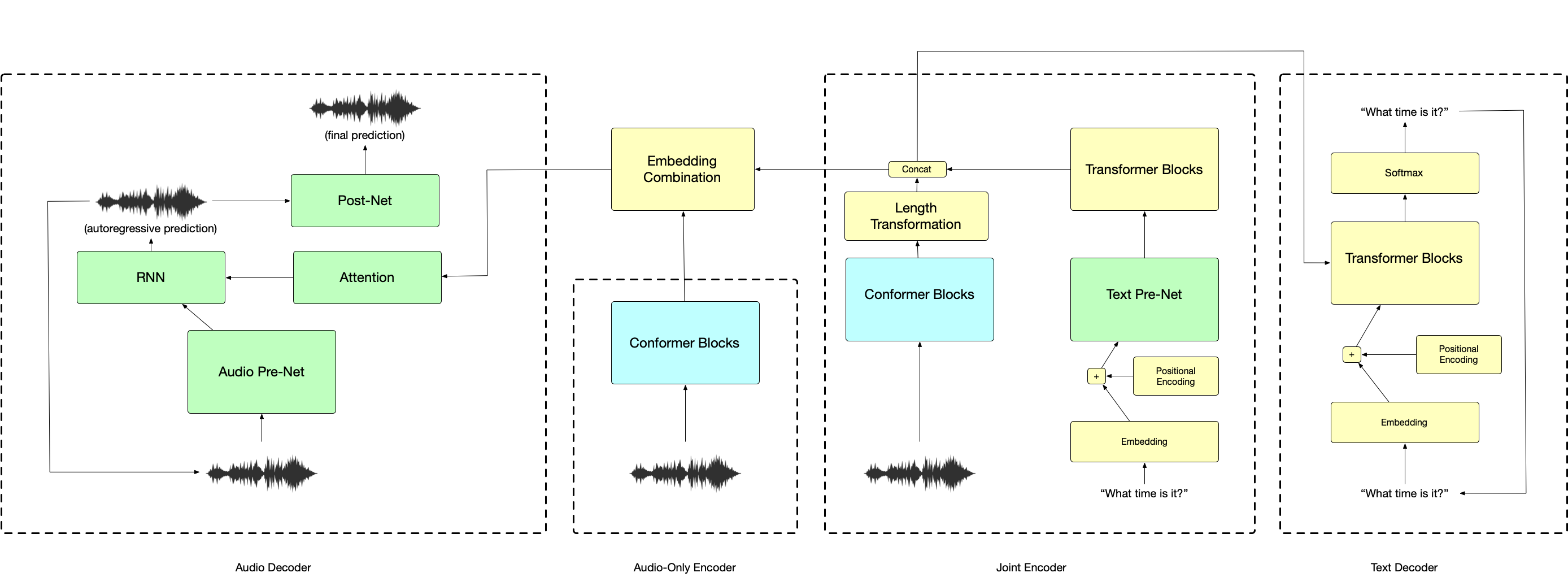}
  \caption{Our joint ASR/TTS model architecture.  Blue components are adapted from Conformer \cite{gulati2020conformer}, green components are adapted from Tacotron 2 \cite{Shen2017}.}
   \label{fig:cascade}
   \vspace{-0.1in}
\end{figure*}

In this section, we describe a joint ASR, TTS, and reconstruction model built on the dual-learning paradigm.  Our architecture is depicted in Figure \ref{fig:cascade}.

\subsection{Architecture Summary}
Our model is trained to consume either text or audio input, and to emit both text and audio.  In that way, for a given input the model either performs ASR and speech reconstruction (audio input) or TTS and text reconstruction (text input).

These tasks are performed by way of a pair of encoders, each of which yield a data representation.  The \say{joint} (or primary) encoder can consume either either audio or text, while the \say{audio-only} (or secondary) encoder consumes only audio input.  There are also two decoders, one corresponding to each of the domains.  The text decoder consumes the output of the joint encoder only, while the audio decoder consumes the outputs of both the joint and audio-only encoders, combined by way of a \say{embedding combination module}, which consists simply of three transformer layers.

For tasks with audio input (ASR and audio reconstruction), the joint encoder consumes zeros instead of text.  For tasks with text input (TTS and text reconstruction), the joint encoder consumes zeros instead of audio, and the embedding combination module consumes zeros instead of the outputs of the audio-only encoder.

\subsection{Encoder Architecture}
The joint encoder is split into two sub-encoders, one to encode audio and one to encode text, each implementing a state-of-the-art encoding scheme.  The audio sub-encoder is based on Conformer \cite{gulati2020conformer}, and consists of 17 conformer blocks with sub-sampling so that the length of the audio input sequence is reduced by a factor of four. The text sub-encoder is based on Tacotron 2 \cite{Shen2017}, and consists of a embedding projection and positional encoding followed by a pre-net and transformer module.  As in \cite{Shen2017}, the pre-net consists of three blocks of a 1D convolution with a 5x1 filter and a dropout layer that zeros out 10\% of its input.  The transformer block borrows from the original transformer architecture in \cite{Vaswani2017} and consists of three blocks of multi-headed self-attention followed by a feed-forward layer.

In order to produce a representation that is agnostic to the input domain, we would like to ensure the joint encoder emits a representation of approximately equal length for both domains.  Otherwise, for example, the audio decoder might learn to model audio reconstruction and TTS separately based on encoding length.   To this end, we adapt the length-transformation component from \cite{Shu2019} by which a representation is compressed into a shorter sequence where each element is a weighted average of elements from the original sequence.  In particular, a sequence $z_1,...,z_M$ is converted to to $\bar{z}_1,...,\bar{z}_N$ (with $N < M$) as:

\begin{align*}
\begin{split}
\bar{z}_j &= \sum_{k=1}^M \sigma(\alpha^j_k)z_k \\
\alpha^j_k &= -\frac{1}{2s}(k - \frac{|x|}{N}j)^2
\end{split}
\end{align*}

where $s$ is a learnable spread parameter and $\sigma$ represents softmax normalization across all weights $\alpha^j_k$ for fixed j.

The audio-only encoder consists simply of four conformer blocks.  These blocks do not include sub-sampling, so the output of the audio-only encoder is the same length as the audio input.

\subsection{Decoder Architecture}
The audio and text decoders are adapted from Tacotron 2 \cite{Shen2017} and Transformer \cite{Vaswani2017} respectively.

The audio decoder consists of a pre-net, autoregressive RNN, and post-net.  The autoregressive component consumes its own previous output, and passes it though a simple audio pre-net which consists simply of a projection and dropout layer.  We then attend to the outputs of the embedding combination module and concatenate the obtained context vector to the processed audio.  This input is passed to a small recurrent network (two LSTM layers) which emits the autoregressive prediction.  As in \cite{Wang2017}, we find that tuning the dropout in the audio pre-net is critical to convergence, since without dropout in the autoregressive input the model simply learns to copy the previous frame.  We find the best results with 10\% dropout.

As in \cite{Shen2017}, we find significant improvement in TTS when the autoregressive decoder output is further processed by a nonautoregressive convolutional post-net.  We use a stack of five convolutions to refine the autoregressive prediction.  During training, we jointly optimize the cross-entropy of both predictions. 

The text decoder is a conventional Transformer \cite{Vaswani2017} decoder, consisting of two blocks each containing a projection, self-attention, and cross-attention.

\section{Experiments \label{sec:experiments}}

We've described how an audio input passed to our model is represented separately by the joint encoder and audio-only encoder.  When optimized to perform the four tasks of ASR, TTS, and audio and text reconstruction, we may naturally imagine two classes of solutions that the model might arrive at:

\begin{itemize}
  \item A \say{disentangled} representation, in which the joint encoder output (which will be consumed by the text decoder) represents that part of the audio signal relevant to the transcript, while the audio-only encoder output (which is only consumed by the audio decoder) represents that part of the audio signal that is not relevant to the transcript.  For example, the joint encoder might represent phonetics, while the audio-only encoder might represent prosody, background noise, and channel effects.
  \item An \say{entangled} representation, in which that part of the audio signal relevant to transcription is not particularly favored by either representation.
\end{itemize}

We seek to observe which of these two representations is learned by our model.  To this end, we train our model fifteen times on the given joint task, arriving at fifteen different solutions to the optimization problem.  We then freeze the parameters of the model, and for each of the fifteen instances we train:

\begin{itemize}
    \item A classifier to determine the speaker ID for a speech example given the model's joint encoder output.
    \item A classifier to determine the speaker ID for a speech example given the model's audio-only encoder output.
\end{itemize}

For a model that has learned a disentangled representation, we expect to be able to predict speaker ID best from the audio-only encoder output, since speaker information is ostensibly required for audio reconstruction but irrelevant to transcription.

\subsection{Entanglement Classifiers}
Speaker IDs are learned using a custom classifier that applies a positional embedding to the selected encoder output followed by three transformer blocks with multi-headed self-attention, five convolutions with a 3x3 filter and stride of 2 and finally a projection and softmax layer. 

\subsection{Model Settings}
As in \cite{Shen2017} and \cite{Wang2017}, we process audio inputs into mel spectrograms with a short-term Fourier transform (STFT) using a frame size of 50 ms and frame hop of 12.5 ms.  We then apply a Han windowing function before applying a mel filterbank, yielding 80-dimensional vectors for our model's audio input.

For text, we choose to use grapheme-level inputs such that the outputs of the embedding layers are 72-dimensional vectors.  While a wordpiece representation might have yielded stronger ASR results, we found that graphemes most reliably ensured convergence of all tasks.

All components use a model dimension of 256, with the model containing about 68 million parameters in total.  Each model is trained with a batch size of 256 split across 16 TPUs.  After 150k steps, we freeze the joint model and train each disentanglement classifier for 100k steps. 

\subsection{Training}
To jointly optimize our four tasks, we split each batch into two halves.  The first half consists of text inputs and represents the TTS and text reconstruction tasks, while the second half consists of audio inputs and represents the ASR and audio reconstruction tasks.  For all elements in the batch, we optimize the loss

\begin{align*}
\centering
L = L_{\text{text}} + \frac{L_{\text{audio\_ar}}}{2} + 
\frac{L_{\text{audio\_final}}}{2}
\end{align*}

where $L_{\text{text}}$ is the cross-entropy loss for the text output, $L_{\text{audio\_ar}}$ is the cross-entropy loss for the audio output before the convolutional post-net, and $L_{\text{audio\_final}}$ is the cross-entropy loss for the audio output after the covolutional post-net.  We find this setup to train more quickly and to converge better than regimens in which tasks alternate across batches.

\subsection{Data}
For training data, we choose the Librispeech corpus \cite{librispeech}.  We see Librispeech as ideal for this experiment since it contains a diverse set of speakers such that there is a significant part of the audio signal to represent outside of the transcript.  We train our models in particular on the \say{clean} subset of the training data, which contains about 460 hours of speech. For WER measurements we evaluate on the \say{clean} test set.

\section{Results \label{sec:results}}

\begin{figure*}[t]
  \begin{subfigure}[t]{0.5\columnwidth}
	  \includegraphics[width=1.0\columnwidth]{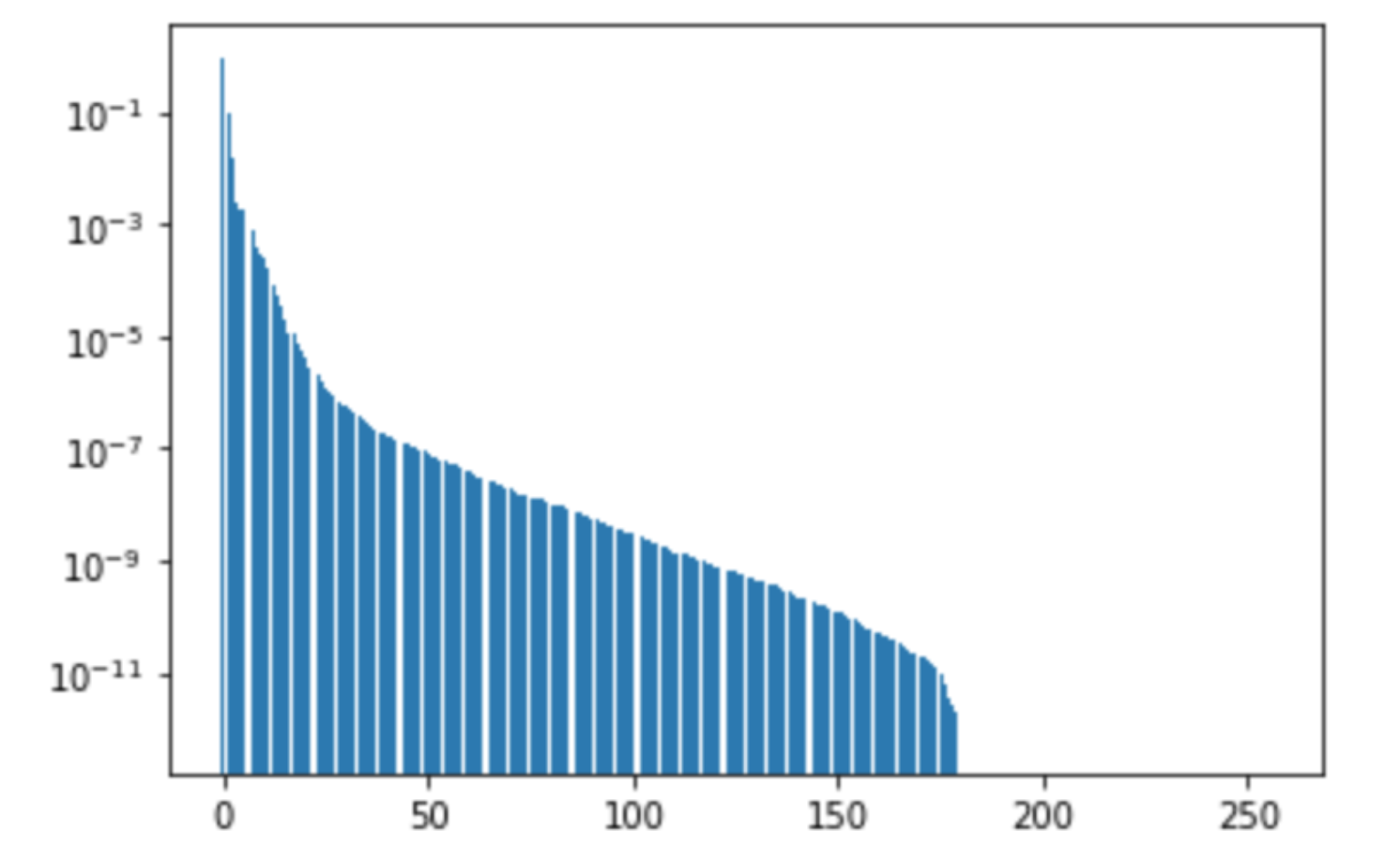}
	  \caption{\footnotesize Audio Embedding, high WER}
  \end{subfigure}
  \begin{subfigure}[t]{0.5\columnwidth}  
	  \includegraphics[width=1.0\columnwidth]{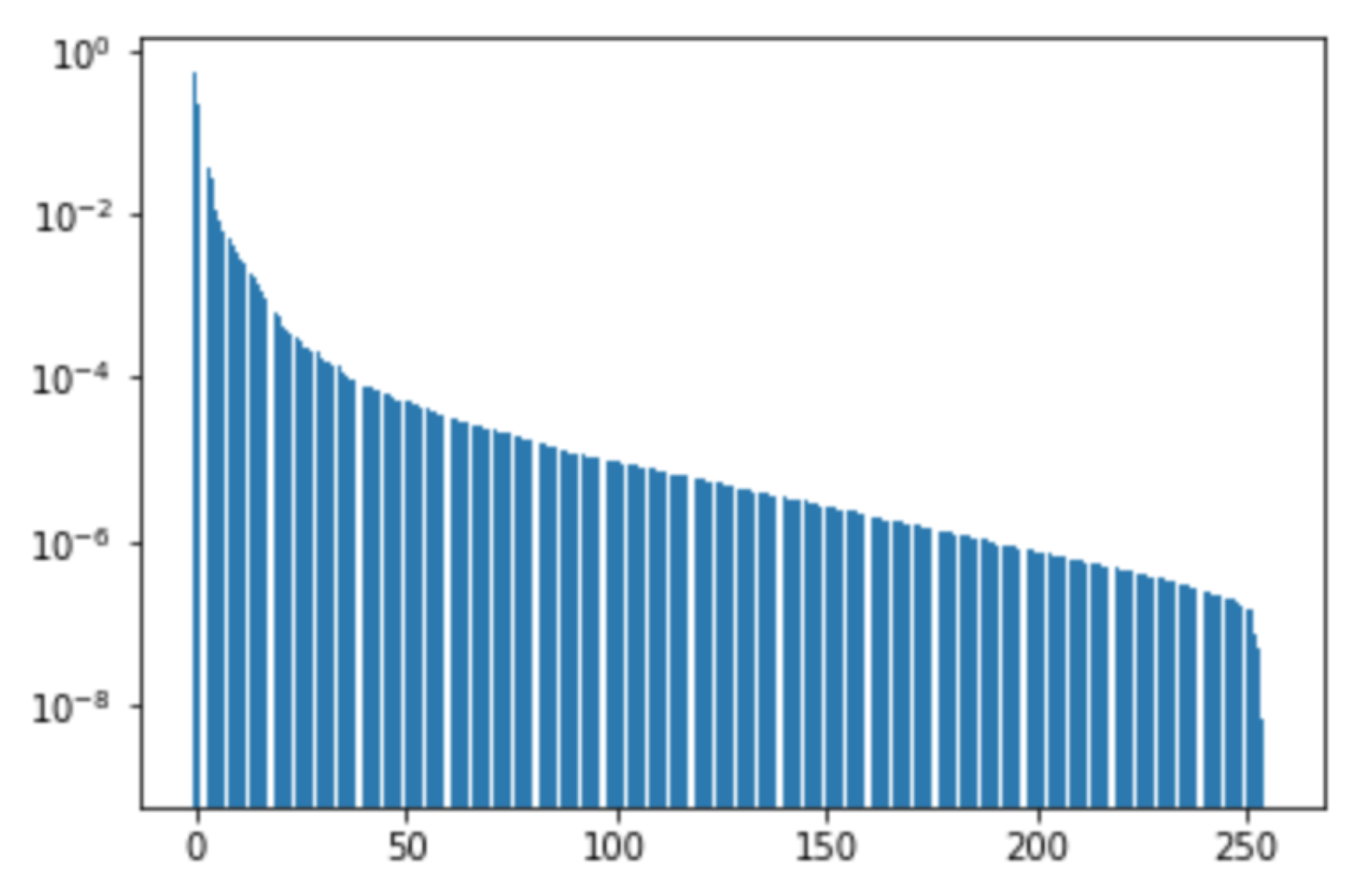}
	  \caption{\footnotesize Audio Embedding, low WER}
  \end{subfigure}
  \begin{subfigure}[t]{0.5\columnwidth}
  	\includegraphics[width=1.0\columnwidth]{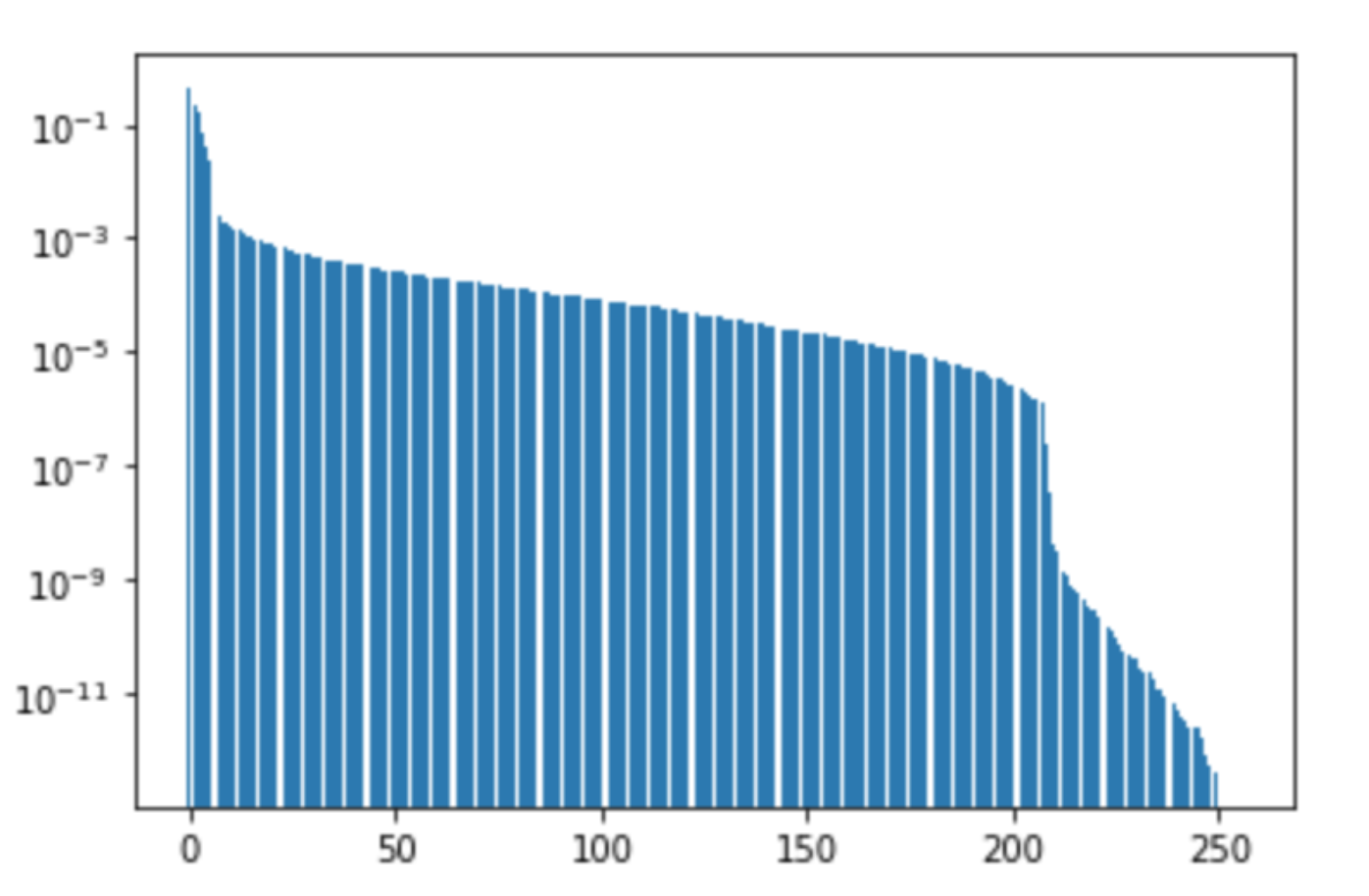}
	\caption{\footnotesize Joint Embedding, high WER}
  \end{subfigure}
  \begin{subfigure}[t]{0.5\columnwidth}
	  \includegraphics[width=1.0\columnwidth]{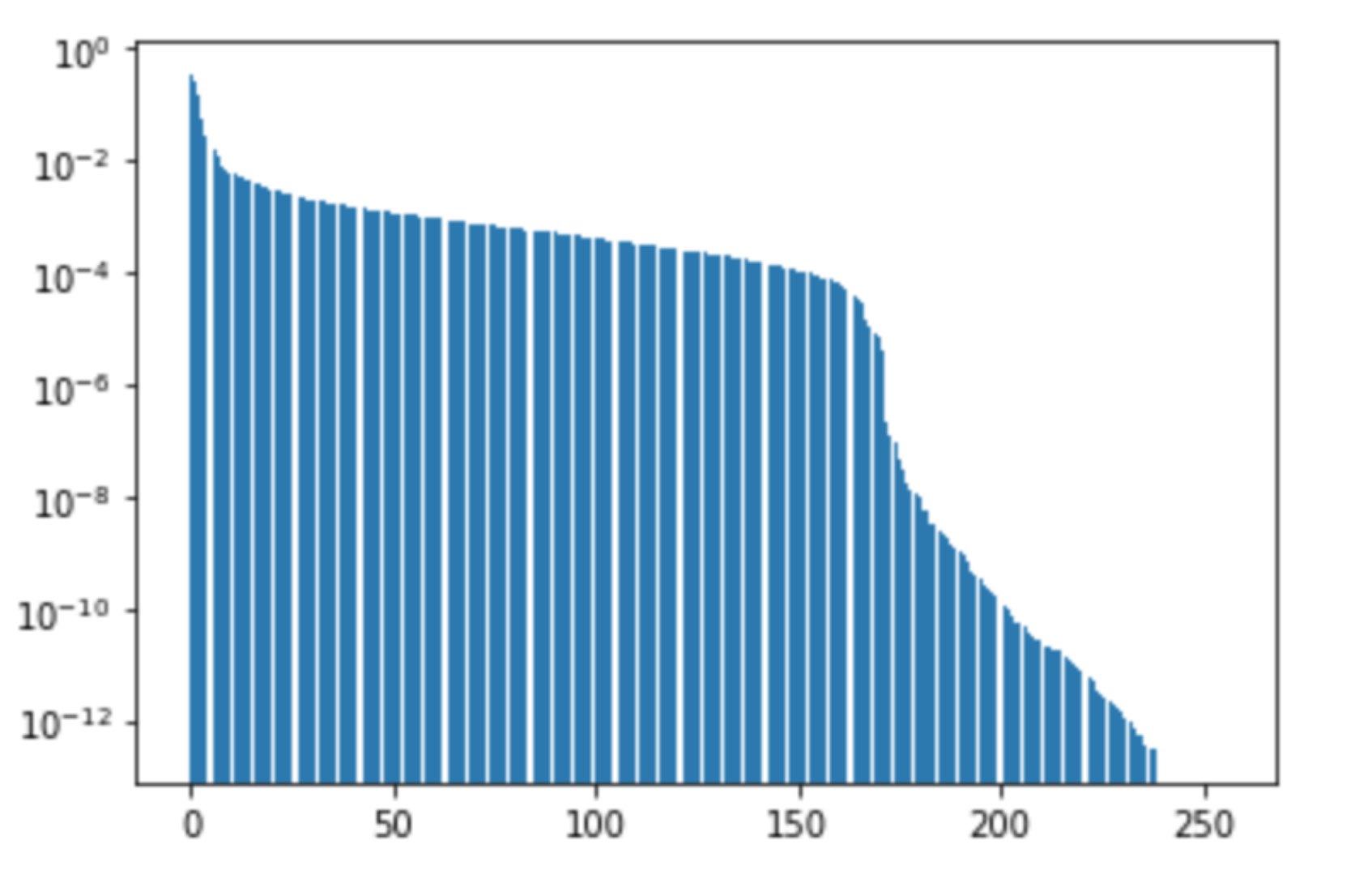}
	  \caption{\footnotesize Joint Embedding, low WER}
  \end{subfigure}
  \caption{The distribution of the singular values of the audio and joint embeddings in strong and weak solutions.}
  \label{fig:svd}
\end{figure*}

\begin{figure}[t]
  \begin{subfigure}[t]{0.47\columnwidth}  
	  \includegraphics[width=1.0\columnwidth]{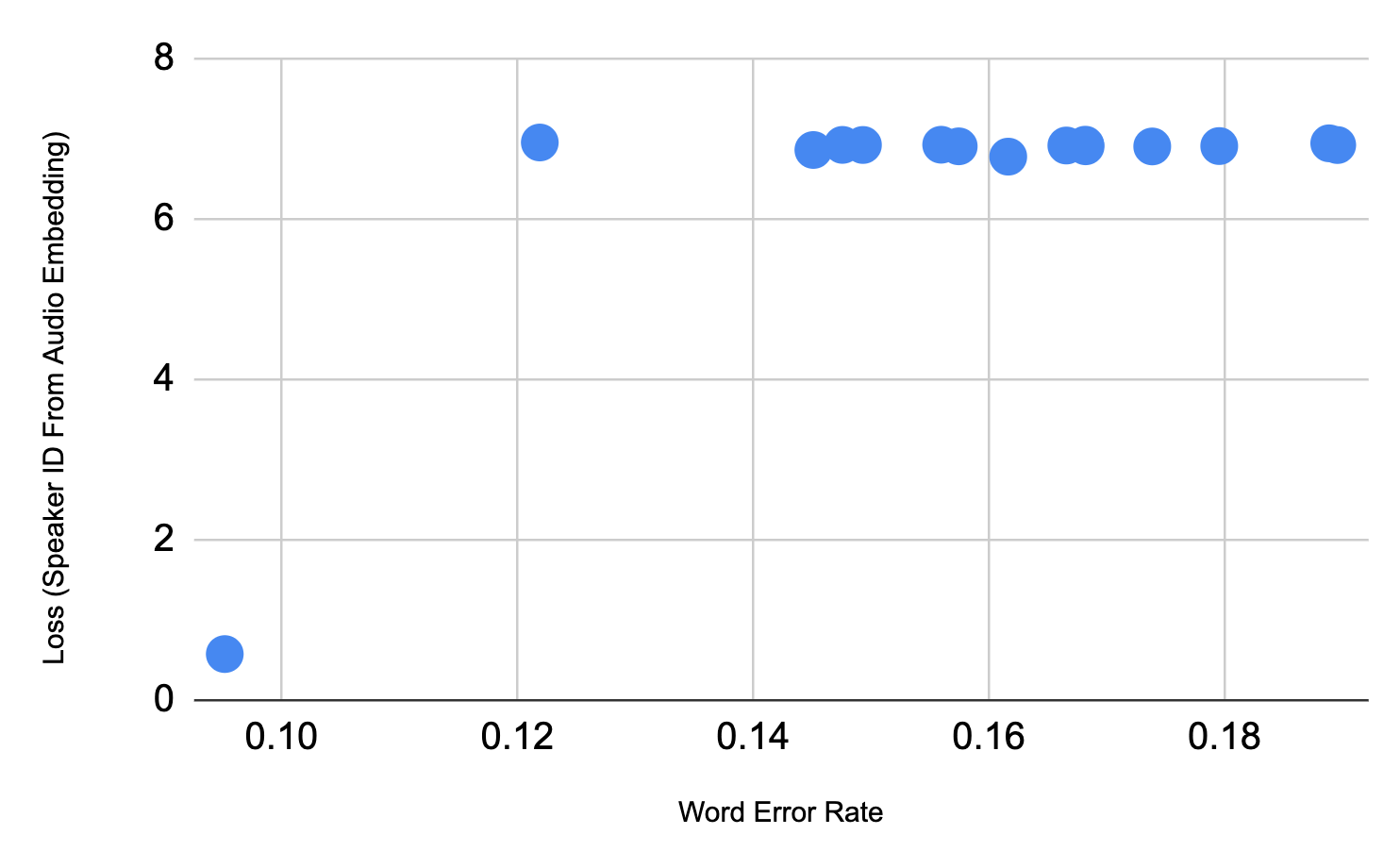}
	  \caption{\footnotesize Speaker ID from Audio}
  \end{subfigure}
  \begin{subfigure}[t]{0.47\columnwidth}
	  \includegraphics[width=1.0\columnwidth]{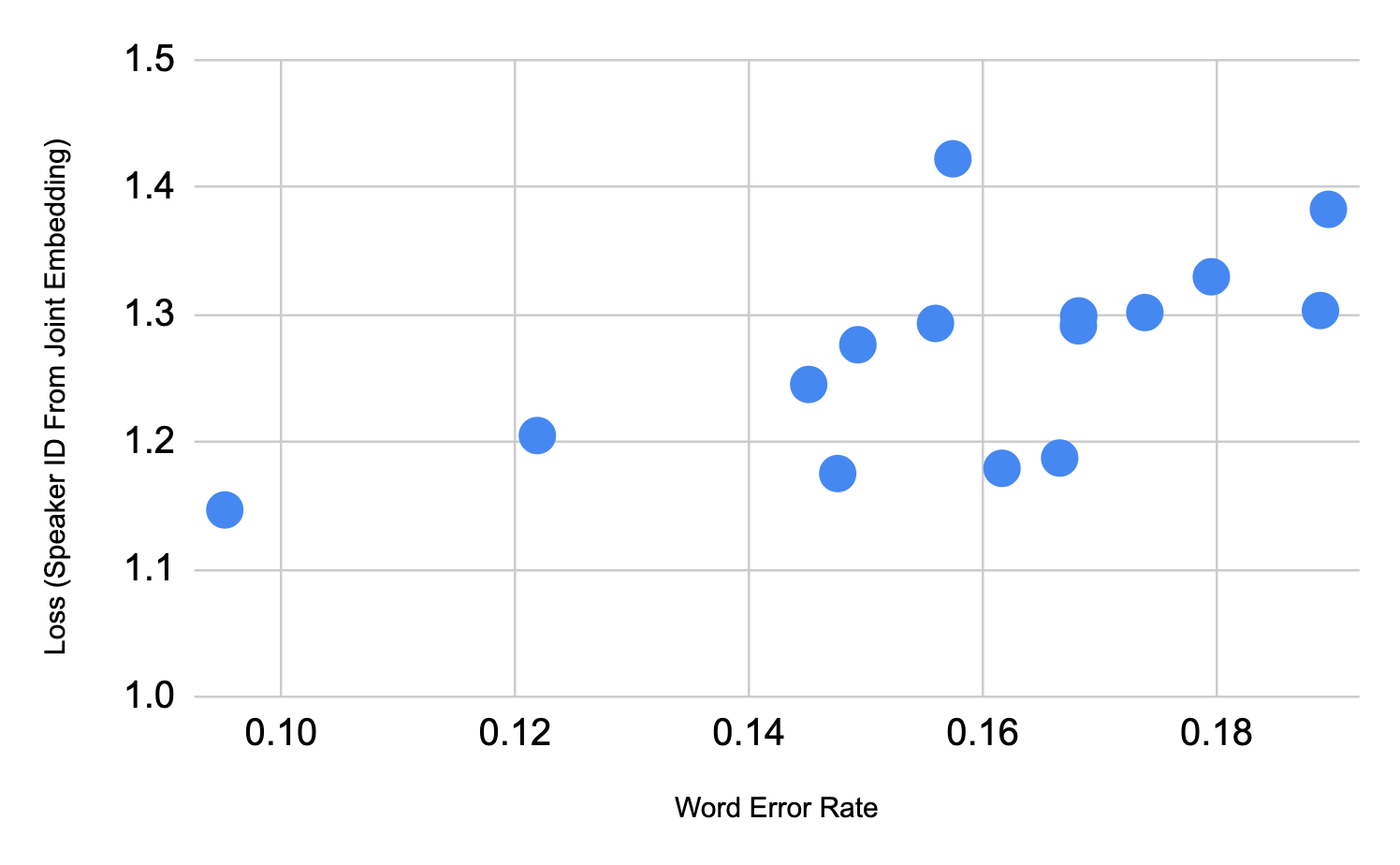}
	  \caption{\footnotesize Speaker ID from Joint}
  \end{subfigure}
  \caption{The relationship between WER and classifier training loss on the four disentanglement tasks measured.}
  \label{fig:classifiers}
\end{figure}

In this section, we report the results of our experiments and analyze the learned representations.

We point out that by the nature of this experiment, our model had to be trained from scratch many times, leading to considerable resource constraints.  These constraints forced us to simplify the training procedure by using only the clean Librispeech data, a small batch size, and a small number of training steps.  This combined with the additional TTS and reconstruction tasks leads to WER values considerably worse than the state of the art for ASR only.  With that in mind, we draw conclusions based on the changes in WER and representation properties across different solutions.

\subsection{Classification}
Figure \ref{fig:classifiers} plots the WER of the joint model against the two classification losses described above after training with frozen encoder parameters.  We quickly make the observation that of our fifteen runs, one has an unusually strong result with a WER of about 9\%.

The speaker ID classification task shows a clear pattern.  The strongest model achieves a training loss that is more than ten times better than the next strongest model on the task from its audio embedding, and more than two times better than on its own joint embedding, suggesting that speaker information has been mostly disentangled from the transcript and localized to the audio embedding.

\subsection{Representation Properties}
Having seen that our model can sometimes, subject to the randomness of training, achieve a much better WER than average, we seek to understand the nature of that stronger, disentangled representation.  In particular, we suspect that in a model without the desired disentanglement, the audio-only embedding is underused. 

To this end, we sample the audio-only and joint representations of our best model and of one of our other models.  Since each input contains a large number of frames, we are able to collect several thousand 256-dimensional vectors from just a few examples.  For each representation, we perform an SVD on those vectors and normalize the squared singular values.  In this manner, we obtain a measurement of the proportion of variance in the representation attributable to each of its 256 dimensions.  We consider a representation with significant variance in a large number of dimensions to be more used by a model than one in which only a few dimensions vary.

The results of these measurements are plotted in Figure \ref{fig:svd}.  We see a stark difference between the distribution of variance in a weak, non-disentangled representation and our strong, disentangled representation.  In particular, the disentangled solution has very few significant dimensions in its audio-only embedding, with the first three dimensions capturing more than 95\% of the variance.  By contrast, the disentangled solution has a much larger number of significant dimensions, with almost 50 dimensions containing more than 0.1\% of the total variance each. 

\subsection{Correlation Loss}
Having observed that strong performance occurs together with a relatively uncorrelated audio-only embedding, we naturally wonder if optimizing for that property at training time will yield better WER.  To test this, we interpolate an additional \say{correlation loss} into the the training of our joint model:

\[L_{\text{corr}} = \alpha \sum_{b \in B} \sum|corr[A\cdot X, A\cdot X] - I|\]

where $b \in B$ are the elements in the batch, $X$ is the matrix formed by stacking the audio representation of the batch element along the time axis, $A$ is a learnable linear projection, $I$ is the identity matrix, and the inner summation adds up each (unsigned) element of the given matrix.  This loss is intended to act as regularization that pushes the off-diagonal elements of the correlation matrix to zero, yielding a representation with uncorrelated elements.  We achieve the strongest results setting the hyperparameter $\alpha=10^{-5}$.

\begin{table}[h!]
    \centering
    \begin{tabular}{|c|c|} \hline
    Model & Average WER \\ \hline
    Non-Disentangled & 15.5\% \\ \hline % 33.0
    Disentangled & 9.8\% \\ \hline % 33.0
    Correlation Loss & 11.7\% \\ \hline % 33.0
    \end{tabular}
    \caption{WER with and without the Correlation Loss}
    \label{tab:wer-corrloss}
    \vspace{-0.2in}
\end{table}

\begin{figure}[t]
  \begin{subfigure}[t]{0.47\columnwidth}
	  \includegraphics[width=1.0\columnwidth]{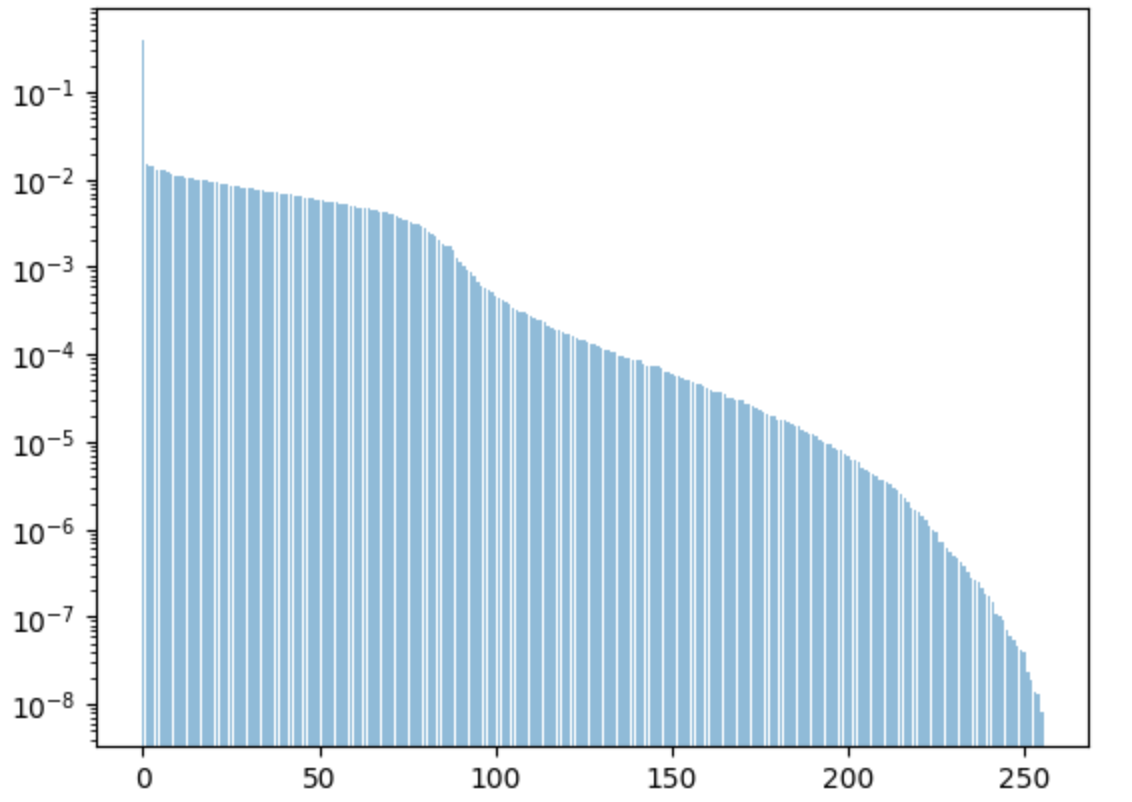}
	  \caption{\footnotesize Audio Embedding, with $L_{\text{corr}}$}
  \end{subfigure}
  \begin{subfigure}[t]{0.47\columnwidth}
  	\includegraphics[width=1.0\columnwidth]{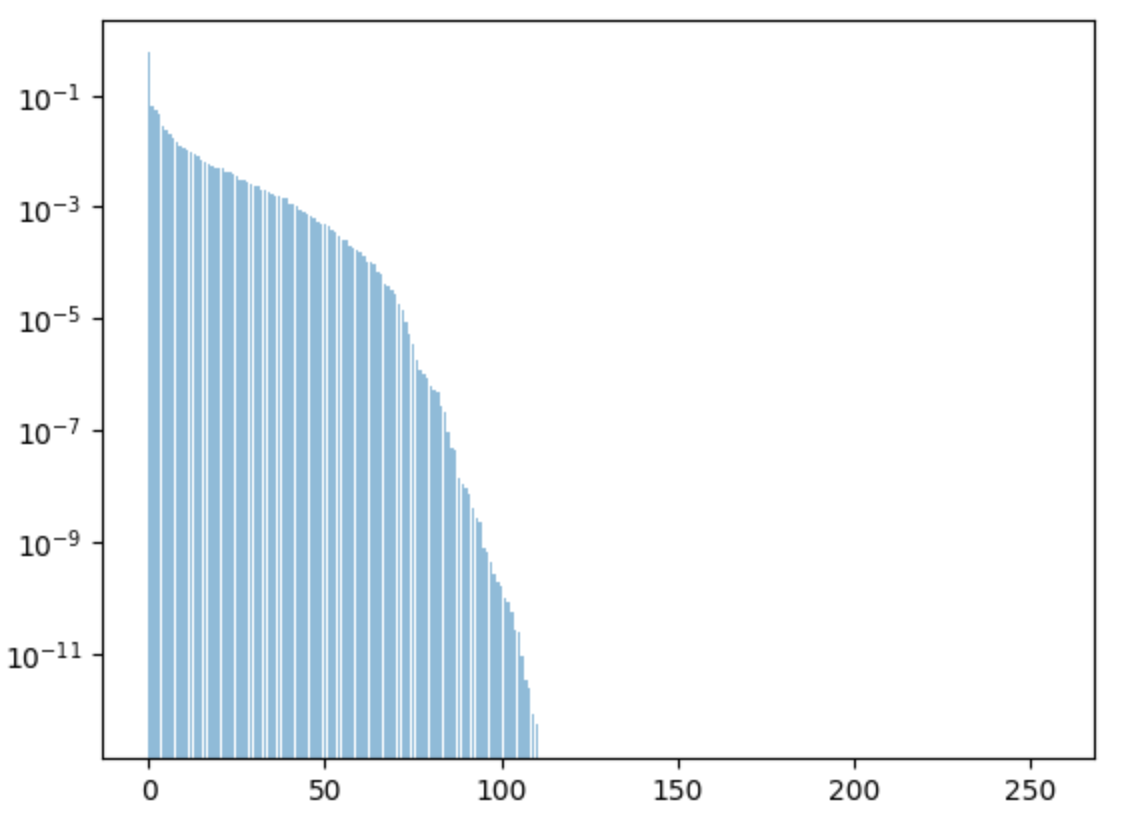}
	\caption{\footnotesize Joint Embedding, with $L_{\text{corr}}$}
  \end{subfigure}
  \caption{Singular value distributions with the correlation loss.}
  \label{fig:svd-corr}
\end{figure}

The distribution of singular values in Figure \ref{fig:svd-corr} shows clearly that the added loss has the intended effect of decorrelating the audio embedding.  It also suggests that this is done by moving information over from the joint embedding, which has become lower-dimensional. WER results are given in Table \ref{tab:wer-corrloss}.  We see that the correlation loss yields on average a 24.5\% reduction in WER. 

\section{Conclusions \label{sec:conclusions}}
In this paper, we presented a novel dual-learning architecture capable of learning a disentangled representation of audio.  We associated disentanglement directly with strong performance on the ASR task and with a high-dimensional audio embedding.

We envision future work in semi-supervised ASR that will train our dual-learning model on supervised data to learn a disentangled audio representation which can then be fine-tuned with both unpaired audio and text data via back-transcription.  We also believe that our work gives rise to more fundamental optimization questions.  We plan to investigate if correlation loss truly promotes disentanglement, or if it reduces WER by some other means.
\newpage
\bibliographystyle{IEEEbib}
\bibliography{main}
\end{document}